\documentclass[pra,
a4paper,
showpacs,
twocolumn,
superscriptaddress]{revtex4-1}
\usepackage{amssymb}
\usepackage{amsmath}
\usepackage{amsfonts}
\usepackage{braket}
\usepackage{graphicx}
\usepackage{bm}
\usepackage{color}
\usepackage{multirow}
\usepackage{natbib}
\usepackage{hyperref}
\usepackage[normalem]{ulem}
\usepackage{verbatim}
\usepackage{dcolumn}
\usepackage{ulem}
\usepackage{float}
\usepackage[utf8]{inputenc}   
\usepackage[T2A]{fontenc}     

\usepackage[dvipsnames]{xcolor}
\usepackage{graphicx}
\def \be {\begin{equation}}
\def \ee {\end{equation}}
\def \bea {\begin{align}}
\def \eea {\end{align}}
\def \p {\partial}
\def \BEA {\begin{eqnarray}}
\def \EEA {\end{eqnarray}}

\def \BC {\begin{cases}}
\def \EC {\end{cases}}

\begin{document}
\title
{ 
Switching rates in Kerr resonator with two-photon dissipation and driving 
}

\author{V.\,Yu. Mylnikov}
\address{Ioffe Institute,
194021 St.~Petersburg, Russia}
\author{S.\,O.~Potashin}
\address{Ioffe Institute,
194021 St.~Petersburg, Russia}
\author{M. S. Ukhtary}
\address{Research Center for Quantum Physics, National Research and Innovation Agency (BRIN), South Tangerang 15314, Indonesia}
\author{G.\,S.~Sokolovskii}
\address{Ioffe Institute,
194021 St.~Petersburg, Russia}


\begin{abstract}
We analytically investigate the switching rate in a two-photon driven Kerr oscillator with finite detuning and two-photon dissipation. This system exhibits quantum bistability and supports a logical manifold for a bosonic qubit. Using Kramer's theory together with the $P$-representation, we derive an analytical expression for the bit-flip error rate within the potential-barrier approximation. The agreement  is demonstrated between analytical calculations and numerical simulations obtained by diagonalization of the Liouvillian superoperator. In the purely dissipative limit, the switching rate increases monotonically with detuning, as the two metastable states approach each other in phase space. However, the exponential contribution to the bit-flip rate exhibits a nontrivial dependence on system parameters, extending beyond the naive scaling with the average photon number. In the presence of large Kerr nonlinearity, the switching rate becomes a nonmonotonic function of the detuning and reaches a minimum at a finite detuning. This effect arises because detuning lowers the activation barrier for weak nonlinearity but increases it for large ones, ensuring a minimum of the switching-rate at nonzero detuning. These results establish key conditions for optimizing the performance of critical cat qubits and are directly relevant for the design of scalable superconducting bosonic quantum architectures.
\end{abstract}

\maketitle

\section{Introduction}

Open quantum systems with quantum bistability \cite{Roberts2020,Beaulieu2025} have gained attention for their ability to realize qubit \cite{Gilchrist2004,Mirrahimi2014}. Instead of isolating the quantum system to protect it from noise, dissipation and external drives can be cleverly used to actively stabilize the logical states \cite{Leghtas2015}. The most promising example is the Schrodinger cat states, where the information is encoded as a symmetric pattern in phase space \cite{Gottesman2001}. Currently, there are two main approaches to generating Schrodinger cat qubits: the dissipative approach \cite{Leghtas2015,Xu2022,Touzard2018,Berdou2023,Marquet2024} and the Hamiltonian approach \cite{Puri2017,Ding2025,Hajr2025}. Both approaches use a two-photon pump, which can be experimentally implemented in microwave resonators with an embedded Josephson junction that exhibits significant nonlinearity \cite{Leghtas2015}. The difference lies in the type of nonlinearity: the dissipative approach exploits two-photon dissipation, while the Hamiltonian approach uses Kerr nonlinearity. Both approaches allow for stabilizing the system on the manifold spanned by the coherent states with opposite phases. Typically, cat qubits are operated at zero frequency detuning, when the pump frequency equals the self-frequency of the cavity mode. However, a recent study \cite{Gravina2023} demonstrates numerically that optimal performance of the cat code can be achieved at nonzero detuning in the presence of two-photon dissipation and Kerr nonlinearity. Understanding the emergence of this optimal detuning and the associated dynamical regimes is thus of practical importance for the design of high-fidelity bosonic quantum codes and scalable quantum computing architectures. As a result, the problem of analytical calculation of the qubit characteristics in such systems becomes relevant to optimize their performance \cite{Chamberland2022}. 

One of the most important characteristics of a cat qubit is the switching or bit-flip rate. It is determined by the quantum fluctuations that can switch the system from one metastable state to another, leading to bit-flip errors. However, analytical studies of the switching rate have proven to be challenging due to noneqelibrium nature of driven-dissipative Kerr cavities. There are three most promising approaches to the analytical calculation of the switching rate. The first is a perturbative expansion in the Lindbladian, which yields both the exponential dependence and the prefactor of the switching rate, but has a limited domain of applicability \cite{Dubovitskii2024,Dubovitskii2025}. The second is the nonperturbative instonton approach \cite{Thompson2022,Carde2025,Sepulcre2025,Lee2025}, which successfully determines the exponential dependence of the switching rate but does not take into account the prefactor. The third is the semiclassical approach \cite{Lin2015,Peano2014,Boness2025,Su2025} that has shown remarkable success in analyzing the Kerr qubit, but it is quite difficult to take into account the effects of nonlinear dissipation in the semiclassical framework.             

In this work, we calculate the switching rate of the Kerr cavity with two-photon drive, two-photon dissipation, and finite detuning. We use Kramer's theory \cite{Hanggi1990}, originally developed for thermally activated escape processes. However, it can be adapted to describe quantum switching between metastable states in driven-dissipative cavities \cite{Kinsler1991}. For analytical calculations, we use the P-distribution governed by a Fokker-Planck-type master equation and the "potential barrier approximation". This allows us to derive explicit expression for the switching rate, including both the exponential dependence and the pre-exponential factor. We then systematically analyze the switching dynamics in the whole range of regimes between the dissipative cat qubits (vanishing Kerr nonlinearity) and Kerr-cat qubits (finite Kerr nonlinearity), at finite frequency detuning.

We demonstrate an increase in the switching rate with an increasing detuning in the dissipative limit as the metastable states become closer in phase space, and the activation process accelerates. However, the exponential factor depends not only on the average photon number and also exhibits a complex dependence on system parameters. Comparison with the Wigner representation confirms similar behavior near the dissipative phase transition point. In particular, we demonstrate that in the presence of strong Kerr nonlinearity, the switching rate exhibits a non-monotonic dependence on detuning, leading to an optimal detuning that minimizes the bit-flip rate. This effect is explained by the behavior of nonlinearities on the effective potential barrier.: while detuning lowers the barrier at low nonlinearity, beyond a threshold value of Kerr constant, it increases the barrier height. Thus, the minimum of the switching rate appears at finite frequency detuning. Our analytical expression for the switching rate demonstrates good agreement with numerical simulation, which is obtained by diagonalization of the Liouvillian superoperator. These findings are crucial for developing stable critical dissipative cat qubits and optimizing two-photon resonators for further application in bosonic quantum computing.

\section{The Model}
We consider a quantum superconducting microwave cavity with Kerr nonlinearity, two-photon pump, and two-photon dissipation. The cavity mode is described by the following Hamiltonian:
\begin{equation}
\hat{H}_{0} = \omega_{c}\,\hat{a}^{\dagger}\hat{a} 
+ \frac{U}{2}\,\hat{a}^{\dagger}\hat{a}^{\dagger}\hat{a}\hat{a},
\end{equation}
where $\hat{a}, \hat{a}^{\dagger}$ are the annihilation and creation operators for photons inside the cavity, $\omega_{c}$ is the cavity frequency and $U$ is the strength of the Kerr nonlinearity. In the same way, a parametric process that coherently adds photons in pairs is described using the following Hamiltonian:  
\begin{equation}
\hat{H}_{\text{2ph}} = i\frac{G}{2} e^{-i2\omega_{p}t}\hat{a}^{\dagger 2} 
-i \frac{G}{2} e^{i2\omega_{p}t}\hat{a}^{2},
\end{equation}
where $G$ is the two-photon pump rate and $\omega_{p}$ is the pump frequency per one photon. The two-photon pump process can be implemented, for example, by coupling two superconducting resonators through a Josephson junction \cite{Leghtas2015}. To remove the explicit time dependence of the drive Hamiltonian, one can move to a rotating frame with frequency $\omega_{p}$. After the unitary transformation, the full Hamiltonian takes the form:
\begin{equation}\label{Hamiltonian}
\hat{H} = -\Delta \hat{a}^{\dagger}\hat{a} 
+ \frac{U}{2}\hat{a}^{\dagger}\hat{a}^{\dagger}\hat{a}\hat{a}
+ \frac{iG}{2}\left(\hat{a}^{\dagger 2} - \hat{a}^{2}\right),
\end{equation}
where $\Delta = \omega_{p} - \omega_{c}$ is the frequency detuning. The two-photon dissipation can be included as a part of the Lindblad master equation on the density matrix $\hat{\rho}$ :
\begin{equation}
\frac{d\hat{\rho}}{dt} = -i[\hat{H},\hat{\rho}] 
+ \frac{\eta}{2}\,\mathcal{D}_{\hat{a}^{2}}(\hat{\rho}),
\end{equation}
where $\hat{H}$ is given in Eq. \eqref{Hamiltonian},  $\eta$ is the two-photon dissipation rate and the Markov–Born approximation is used. Here, $\mathcal{D}_{\hat{a}^{2}}(\hat{\rho})$ is a Lindblad superoperator: 
\begin{equation}
\mathcal{D}_{\hat{a}^{2}}(\hat{\rho}) = 2\hat{a}^{2}\hat{\rho}\hat{a}^{2\dagger} 
-\hat{a}^{2\dagger} \hat{a}^{2}\hat{\rho} - \hat{\rho}\hat{a}^{2\dagger} \hat{a}^{2},
\end{equation}
and $\hat{a}^{2}$ is the quantum jump operator for the two-photon dissipation. This model therefore incorporates three essential mechanisms: cavity detuning, Kerr nonlinearity, and two-photon drive with corresponding two-photon losses. 
\section{The Semiclassical theory}\label{Semi}
\begin{figure}[ht]
    \centering
    \begin{minipage}{0.45\textwidth}
        \centering
        \includegraphics[width=0.9\linewidth]{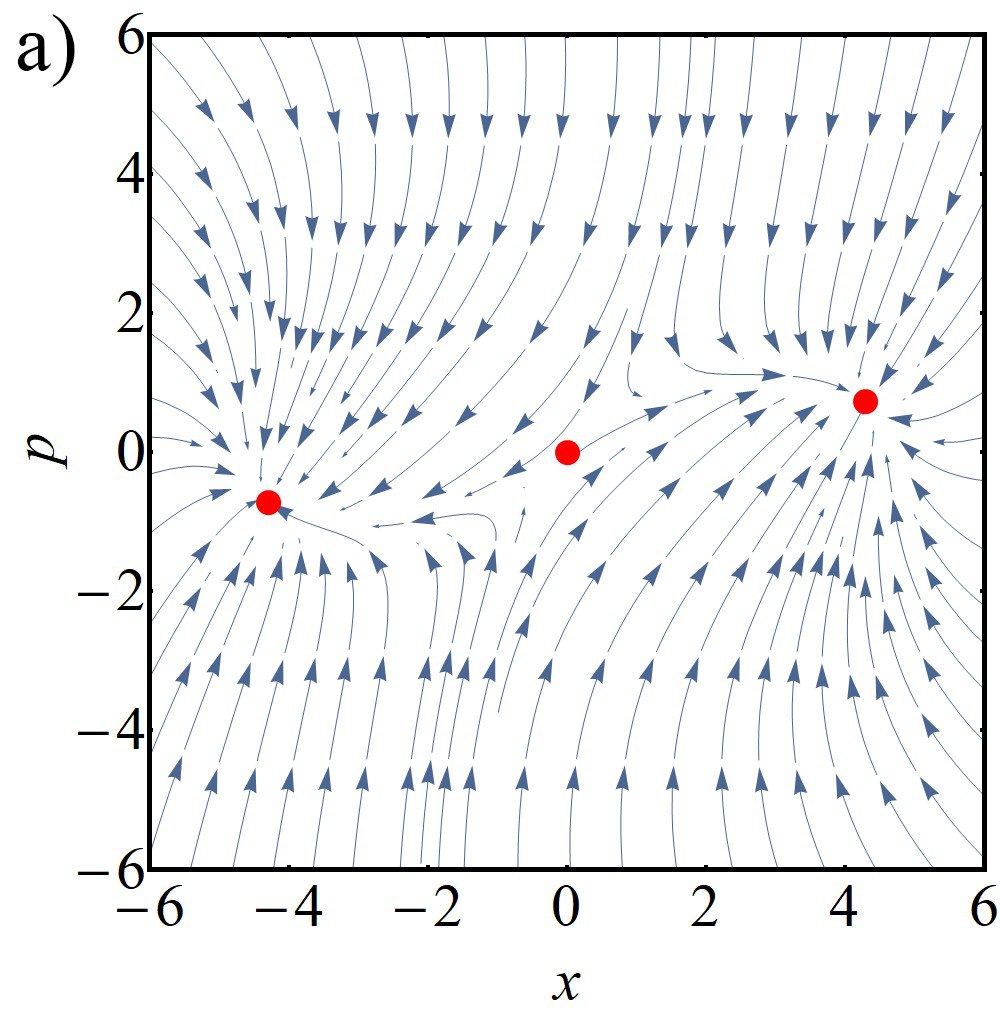}
    \end{minipage}
    \hfill
    \begin{minipage}{0.45\textwidth}
        \centering
        \includegraphics[width=0.9\linewidth]{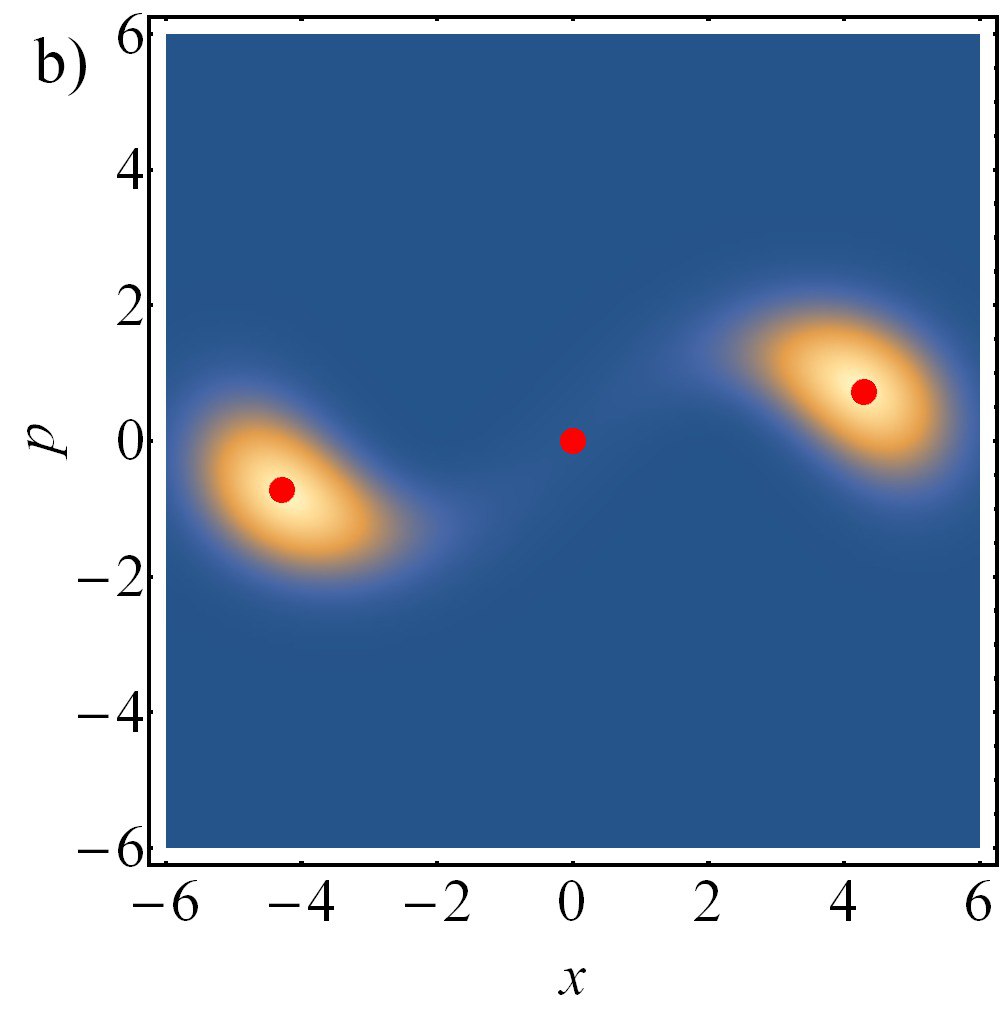}
    \end{minipage}
\caption{(a) The phase portrait of the semiclassical Eqs.~\eqref{xp}. Blue arrows indicate the drift. (b) The Wigner function as a function of the photonic quadratures $x$ and $p$. The red dots highlights location of the semi-classical solutions. In this calculation we set $G=6, \Delta=3,\eta=\sqrt{3}/2,U=1/2$.}
    \label{fig:xp}
\end{figure}

Let us first analyze the steady state of the system. For simplicity, we will consider the semiclassical approximation \cite{Bartolo2016,Meaney2014}, where all quantum fluctuations are neglected. In this approximation, the coherent field amplitude $\alpha(t) = \mathrm{Tr}\left[\hat{a}(t)\hat{\rho}\right]$ obeys the following simple equation:  
\begin{equation}\label{alpha}
 \begin{split}
 \p_t \alpha=i  \Delta \alpha  +  G\alpha^*-(\eta +i U)\alpha ^2 \alpha^*  ,
 \end{split}
\end{equation}
The stationary solutions are defined by the condition $\dot{\alpha}=0$. As a result, in addition to the trivial solution $(\alpha=0)$, one can find the following stationary solutions $\alpha=\pm\alpha_0$, where $\alpha_0=\sqrt{n_0}\exp(i\theta_{0})$, $n_0$ is an average photon number, and $\theta_0$ is a phase factor: 
 \begin{align}
     &n_0=\frac{\sqrt{G^2 \left(\eta ^2+U^2\right)-\Delta ^2 \eta ^2}+\Delta  U}{\eta ^2+U^2},\\
     &\exp(i2\theta_{0})=\frac{\sqrt{G^2 \left(\eta ^2+U^2\right)-\Delta ^2 \eta ^2}+i \Delta  \eta }{G (\eta +i U)}.
 \end{align}
Let us discuss the phase portrait of Eq. \eqref{alpha}. For this purpose, it is convenient to rewrite Eq.~\ref{alpha} in terms of the photon quadratures $\alpha=(x+i p)/\sqrt{2}$:  
\begin{equation}\label{xp}
 \begin{split}
   &\p_t x=G x-\Delta  p-\frac{1}{2}   \left(p^2+x^2\right) (\eta x-U p),\\
    &\p_t p=\Delta  x-G p-\frac{1}{2}  \left(p^2+x^2\right)(\eta p+U x),
 \end{split}
\end{equation}
where $x(t)=\sqrt{2}\mathrm{Re}\left[\alpha(t)\right]$ and $p(t)=\sqrt{2}\mathrm{Im}\left[\alpha(t)\right]$. It turns out that the system has two stable points ($x=\pm x_0$, and $p=\pm p_0$), where $x_0=\sqrt{2}\text{Re}(\alpha_0)$ and $x_0=\sqrt{2}\text{Im}(\alpha_0)$, and one saddle point ($x=0$, $p=0$), as shown in Fig.~\ref{fig:xp}a. This is true for $|\Delta|<G$. The type of stability of these points is discussed in detail in the Appendix \eqref{AppA}. The presence of quantum fluctuations causes stable stationary solutions to become metastable, acquiring a finite lifetime. Thus, a nonzero probability of switching between different stable solutions emerges \cite{Dubovitskii2025,Thompson2022,Carde2025}. A detailed discussion of the mechanism of such switching is presented in the following Sections.
\section{The Kramers theory}
\subsection{P representation}
In the previous Section we neglect quantum fluctuations and discuss the fixed point of the semiclassical system. Here, our investigation focuses on how the semi-classical picture changes in a full quantum description. In this Section we will demonstrate that semiclassical stable solutions become metastable with an exponentially small transition rate between states. To analyze the quantum switching between two classical metastable points, we apply Kramer's theory \cite{Gardiner}. The initial step involves formulating the problem in terms of a Fokker-Planck-type equation of motion, from which the corresponding stationary distribution function is derived.  This is achieved through the complex P-representation of the density operator, $P(\alpha,\beta)$, which is defined by Drummond et al. as follows \cite{Drummond1980,Kinsler1991}: 
\begin{equation}
\hat{\rho}=\int_\mathcal{C}\!\!d\alpha \int_{\mathcal{C}'}\!\!d\beta\, \frac{\ket{\alpha}\bra{\beta^*}}{\braket{\beta^*|\alpha}}\, P(\alpha,\beta),
\end{equation}
where $\ket{\alpha}$ and $\ket{\beta^*}$ are the coherent states $\left(\hat{a}\ket{\alpha}=\alpha \ket{\alpha}\right)$ and $\alpha,   \beta$ are complex variables that need to be integrated over individual contours $C$, $C'$ in the complex plane. The introduced P-distribution function satisfies the Fokker-Planck-like equation of motion \cite{Bartolo2016}: 
\begin{equation}\label{Fokker_PlanckP}
\,\partial_t P= \sum_{i,j=\alpha,\beta} \partial_i \left[-A^i P + \frac{1}{2} \partial_j \left(D^{ij} P \right)  \right],
\end{equation}
where the drift vector and diffusion matrix have the following form:
\begin{equation}
\begin{split}
&\bar{A}= 
\begin{pmatrix} 
i\Delta \alpha + G\beta-\kappa_2 \alpha^2 \beta \\ 
-i\Delta \beta + G\alpha-\kappa_2^* \beta^2 \alpha
\end{pmatrix},
\quad
\\
&D= 
\begin{pmatrix} 
G-\kappa_2 \alpha^2 &  0  \\ 
0 &  G-\kappa_2^* \beta^2
\end{pmatrix},
\end{split}
\end{equation}
where we define the complex nonlinear coupling constant $\kappa_2=\eta+i U$. Recently, a stationary solution of the Fokker-Planck in Eq. \eqref{Fokker_PlanckP} was derived \cite{Minganti2016,Bartolo2016} and it is given by:
\begin{equation}\label{StationarySol} 
P_{\rm ss}(\alpha,\beta)=\exp[-\Phi (\alpha,\beta)],
\end{equation}
where the effective complex potential is as follows: 
\begin{equation}\label{PotentialP}
 \begin{split}
\Phi(\alpha,\beta&)=-2\alpha\beta+\left(1+\frac{i\Delta}{\kappa_2}\right)\ln\left[\alpha^2-\frac{G}{\kappa_2}\right]+\\ &+\left(1-\frac{i\Delta}{\kappa_2^*}\right)\ln\left[\beta^2-\frac{G}{\kappa_2^*}\right].
\end{split}
\end{equation}
The effective potential \eqref{PotentialP} has five extrema. The first extrema $(\alpha=0,\beta=0)$ corresponds to the saddle point of the potential. The second and third extrema are the classical points $(\alpha=\pm\alpha_{cl},\beta=\pm\alpha_{cl}^*)$, as long as $\alpha$ and $\beta$ are complex conjugate quantities. The last two extrema points are quantum $(\alpha=\pm\alpha_{q},\beta=\mp\alpha_{q}^*)$, since they correspond to the density matrix $\ket{\pm\alpha_{q}}\bra{\mp\alpha_{q}}$, which has no analog in classical physics. The complex amplitudes corresponding to the classical $\alpha_{cl}=|\alpha_{cl}|e^{i\theta_{cl}}$ and quantum $\alpha_{q}=|\alpha_{q}|e^{i\theta_{q}}$ points are given by the following:  
\begin{equation}
 \begin{split}
&|\alpha_{cl}|=\sqrt{\frac{\sqrt{G^2(U^2+\eta^2)-\Delta ^2\eta^2}+U \Delta}{U^2+\eta^2 }+1},\\ 
&|\alpha_{q}|=\sqrt{\frac{\sqrt{G^2(U^2+\eta^2)-\Delta ^2\eta^2}-U \Delta}{U^2+\eta^2 }-1},
 \end{split}
\end{equation}
and the phase factor is as follows:
\begin{equation}
\begin{gathered}\label{angles1}
    e^{i2\theta_{cl}}=(\sqrt{G^2(U^2+\eta^2)-\Delta ^2\eta^2}+i \eta\Delta)/G(\eta+iU),\\
e^{i2\theta_{q}}=(\sqrt{G^2(U^2+\eta^2)-\Delta ^2\eta^2}-i \eta\Delta)/G(\eta+iU).
\end{gathered}
\end{equation}
The Wigner function for a given P-distribution \eqref{StationarySol} is given by \cite{Bartolo2016}: 
\begin{equation}\label{Wigner}
    W(x,p)=\frac{2}{\pi }\frac{{{\left| {}_{0}{{F}_{1}}\left( \frac{1}{2}-\frac{i\Delta }{\eta };\frac{G}{2 \eta }{{(x-ip)}^{2}} \right) \right|}^{2}}}{{}_{1}{{F}_{2}}\left( \frac{1}{2};\frac{1}{2}-\frac{i\Delta }{\eta },\frac{1}{2}+\frac{i\Delta }{\eta };\frac{G^2}{{{\eta }^{2}}} \right)}{{e}^{-({{x}^{2}}+{{p}^{2}})}},
\end{equation}
where $_pF_q$ denotes a generalized hypergeometric function. One can see a comparison between the Wigner function \eqref{Wigner} shown in Fig. \ref{fig:xp}b and the semiclassical dynamic of Eqs. \eqref{alpha}, which is depicted in Fig. \ref{fig:xp}a.   

Let us demonstrate that the switching rate is determined by the quasi-classical activation processes between two classical points through the saddle point. As we show below in Fig.\eqref{Fig3}, our calculation correctly reproduces numerical simulations of the Liouvillian gap obtained by diagonalization of the Liouvillian superoperator.  Since P-distribution satisfies the Fokker-Planck-like equation of motion \eqref{Fokker_PlanckP} and its stationary solution \eqref{StationarySol} was found, methods of well-developed Kramer`s theory can be used. The switching rate can be calculated using the generalization of the Eyring–Kramers law for the multidimensional case \cite{Berglund2011}:

\begin{equation}\label{Kramers1}
\begin{gathered}
\Gamma=\frac{D_0 |\Lambda_1(0,0)|}{2 \pi}\sqrt{\frac{\det(\nabla^2\Phi(\alpha_{cl},\alpha_{cl}^*))}{|\det(\nabla^2\Phi(0,0))|}}
e^{\Phi(\alpha_{cl},\alpha_{cl}^*)-\Phi(0,0)},
\end{gathered}
\end{equation}
where $\Lambda_1(\alpha,\beta)$ is the single negative eigenvalue of Hessian $\nabla^2\Phi(\alpha,\beta)$ and $D_0=G$ is the diffusion coefficient at the saddle point. 
In a recent paper \cite{Carde2025} the exponent of \eqref{Kramers1}, was estimated using the Keldysh path integral approach. As we show above, expression \eqref{Kramers1} also correctly reproduces the preexponential factor of the decoherence rate. After some algebra, which can be found in Appendix \eqref{AppB}, the decoherence rate takes the following form:
\begin{equation}\label{gamma}
\Gamma=B\exp(-\delta\Phi),
\end{equation}
where $B$ is the preexponential factor:
\begin{equation}
\begin{gathered}
B=\frac{2}{\pi }
\frac{|\Delta|}{G}
\left(G^2 \left(\eta ^2+U^2\right)-\Delta ^2 \eta ^2\right)^{1/4}\times\\
 \times \sqrt{\frac{\sqrt{G^2 \left(\eta ^2+U^2\right)-\Delta
   ^2 \eta ^2}+\Delta  U}{\eta ^2+U^2}},
\end{gathered}
\end{equation}
and $\delta\Phi$ is an effective potential barrier height expressed as follows:

\begin{widetext}
\begin{equation}\label{Porential} 
\begin{gathered}
\delta\Phi=\frac{2\left(\sqrt{G^2 \left(\eta ^2+U^2\right)-\Delta ^2 \eta ^2}+\Delta  U\right)}{\eta ^2+U^2}-\frac{2 \Delta  \eta  \left(\arctan\left(\frac{\sqrt{G^2 \left(\eta
   ^2+U^2\right)-\Delta ^2 \eta ^2}}{\Delta  \eta }\right)+\arctan\left(\frac{U}{\eta }\right)\right)}{\eta
   ^2+U^2}-\\
-\frac{2\Delta  U}{\eta ^2+U^2}\ln\left(\frac{|\Delta|}{G}\right),
\end{gathered}
\end{equation}
\end{widetext}
where we use the assumption of small nonlinearities ($U,\eta<<\Delta,G$). The comparison between the analytical expression \eqref{gamma} for the switching rate and the numerical simulation is shown in Fig. \ref{Fig4}. For the latter, we perform a numerical diagonalization of the Linbladian superoperator on the truncated Fock basis and find the eigenvalue with the smallest real part $\lambda_1$, which determines the switching rate as follows $\Gamma=-\mathrm{Re}(\lambda_1)$. In Fig. \ref{Fig4} the switching rate is shown as a function of a frequency detuning $\Delta$. In the following Sections, we will consider in detail two cases where the Kerr nonlinearity is zero and greater than zero.
\subsection{Zero Kerr nonlinearity and dissipative cat qubit}
\begin{figure}
\centering
\includegraphics[width=0.45\textwidth]{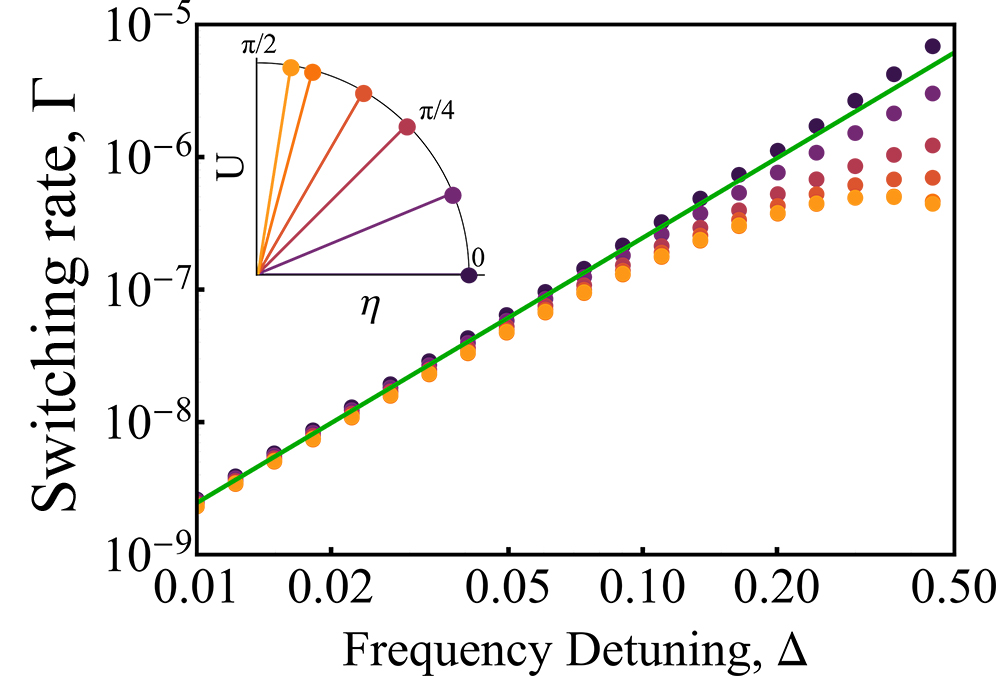}
\caption{The switching rate, $\Gamma$, obtained by «small detuning approximation» \eqref{GammaSmallDetunings} (green line) and numerical diagonalization of the Liouvillain superoperator (dots), vs the frequency detuning $\Delta$. 
The nonlinear coupling constant $\kappa_2=\eta+i U=|\kappa_2|\exp(i\theta)$ has a fixed modulus $|\kappa_2|=1$, but different values of the angle~$\theta=\arctan(U/\eta)= \pi/2\cdot(0,0.25,0.5,0.66,0.83,0.9)$, as shown in the inset. An increase in Kerr nonlinearity $U$ (decrease in the two-photon dissipation $\eta$) corresponds to a color change of the dots from dark violet to bright orange. The two-photon pump rate is set to $G = 6$, respectively.}
\label{Fig3}
\end{figure}

In the previous Section, we have calculated the switching rate between two metastable states using the P-representation approach. Here, we will focus on the special case where the Kerr nonlinearity is zero ($U=0$). In the resonance case $\Delta=0$ (i.e. the two-photon pump is resonant with the cavity mode, i.e. $\omega_p=\omega_c$), only two physical processes are present: a coherent two-photon pump with rate $G$ and two-photon dissipation with rate $\eta$. This system has four distinct stationary density matrices: $\ket{\alpha_{cl}}\bra{\alpha_{cl}}$, $\ket{-\alpha_{cl}}\bra{-\alpha_{cl}}$, $\ket{\alpha_{cl}}\bra{-\alpha_{cl}}$, and $\ket{-\alpha_{cl}}\bra{\alpha_{cl}}$, where $\alpha_{cl}=\sqrt{G/\eta}$  and $\ket{\alpha}$ is a coherent state. The corresponding density matrices form a dark space \cite{Thompson2022}, which is known as a dissipative Schrödinger-cat qubit. This makes the two-photon dissipative cavities a promising platform for practical bosonic quantum codes \cite{Mirrahimi2014,Leghtas2015,Touzard2018,Gautier2022}. The fourfold degeneracy of dark space is a consequence of a strong $Z_2$ symmetry \cite{Kamenev, Thompson2022, Lieu2020}. However, nonzero detuning ($\Delta\neq0$) produces an exponentially weak dephasing rate within the qubit dark space. As shown by Foster Thompson, Alex Kamenev, and Kirill Dubovitskii \cite{Thompson2022, Dubovitskii2025}, the dependence of the dephasing (switching) rate on the system parameters for small detunings is as follows (Fig. \ref{Fig3}): 
\begin{equation}
\begin{gathered}\label{exactRate}
\Gamma=\frac{4 \Delta^2 }{\eta} 
 \exp\left[-2G/\eta\right].
\end{gathered}
\end{equation}
Here, the exponential factor $\exp(-2G/\eta)$ suppresses the switching rate, as long as the activation process occurs between two distinct classical points $\pm\alpha_{cl}=\pm\sqrt{G/\eta}$. The greater the separation between these two states in the phase space, the longer the switching time. Furthermore, the pre-exponential factor of Eq. \eqref{exactRate} grows quadratically with frequency detuning $\Delta$. It is also inversely proportional to the two-photon dissipation rate~$\eta$. 

\begin{figure*}
\includegraphics[width=0.7\textwidth]{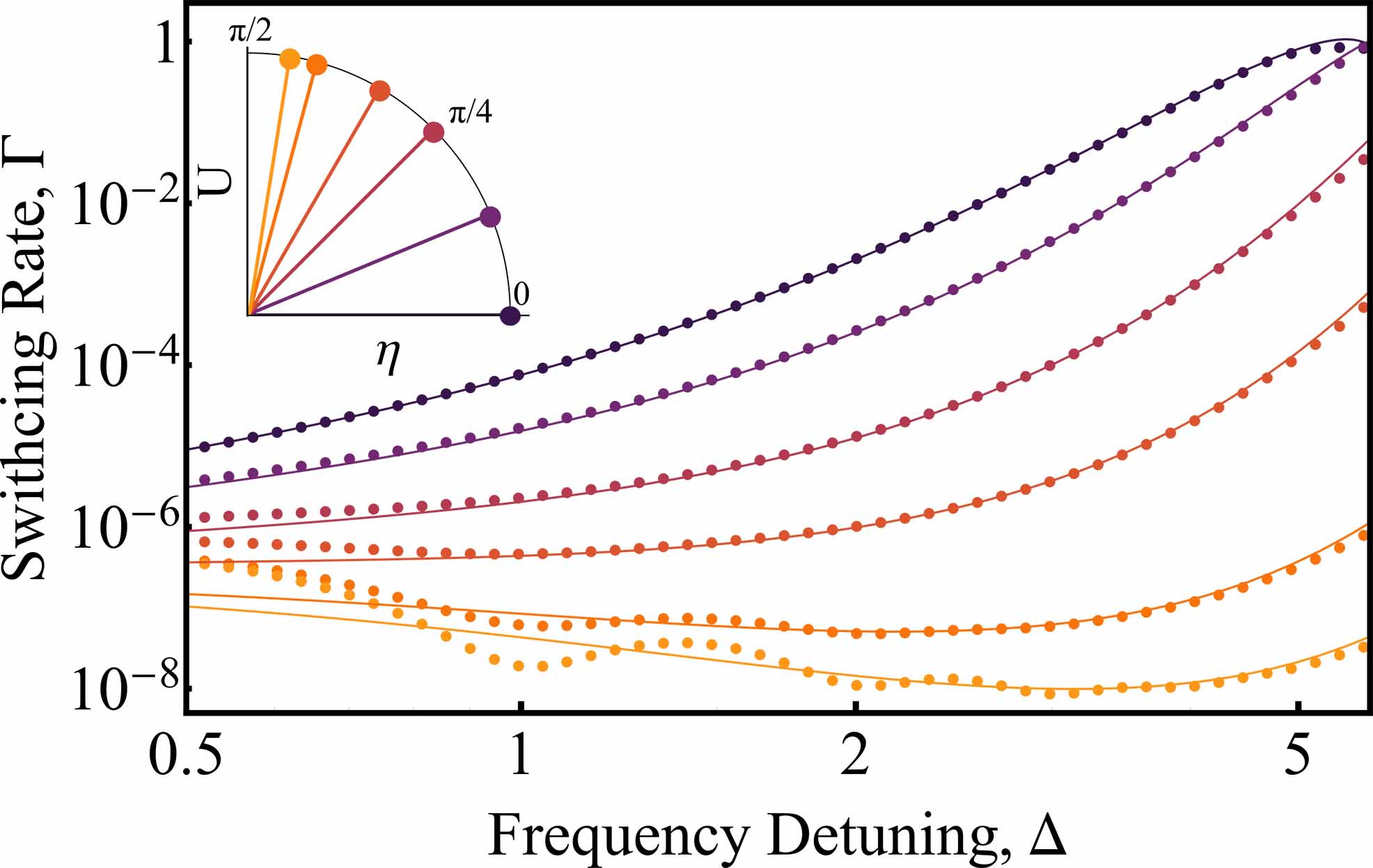}
\caption{The switching rate, $\Gamma$, obtained by «potential barrier approximation» \eqref{gamma} (curves) and numerical diagonalization of the Liouvillain superoperator (dots), vs the frequency detuning $\Delta$. 
The nonlinear coupling constant $\kappa_2=\eta+i U=|\kappa_2|\exp(i\theta)$ has a fixed modulus $|\kappa_2|=1$, but different values of the angle $\theta=\arctan(U/\eta)=\pi/2\cdot(0,0.25,0.5,0.66,0.83,0.9)$, as shown in the inset. An increase in Kerr nonliniarity $U$ (decrease in the two-photon dissipation $\eta$) corresponds to a color change of the curve from dark violet to bright orange. The two-photon pump rate is set to $G = 6$, respectively.}
\label{Fig4}
\end{figure*}
Let us compare our findings \eqref{Kramers1} obtained by the «potential barrier approximation» with Eq. \eqref{exactRate} \cite{Thompson2022, Dubovitskii2025}. In the small detuning regime, where $\eta<<G,\Delta$, \eqref{Kramers1} is transformed as follows: 
\begin{equation}\label{gamma3}
\begin{gathered}
\Gamma\propto\eta 
   \exp\left[-2G/\eta\right].
\end{gathered}
\end{equation}
One can see from Eq. \eqref{gamma3} that the exponent is given by $-2G/\eta$, which reproduces the result \eqref{exactRate} discussed above. However, the pre-exponential factor of Eq. \eqref{gamma3} is constant and depends on the two-photon dissipation rate. Thus, the pre-exponential factor of \eqref{gamma3} differs from the predictions of the exact answer \eqref{exactRate}. This discrepancy between two prefactors occurs because the "potential-barrier approximation" overestimates the switching rate for small detunings. The same situation occurs in the classical Kramers problem, where the "potential barrier approximation" ceases to be applicable for the small dissipation regime \cite{Hanggi1990}. However, two asymptotics (Eq. \eqref{exactRate}) and (Eq. \eqref{gamma3} ) match near a fairly small detuning value, proportional to the two-photon dissipation rate $\Delta\propto \eta$. Above this point, the calculated switching rate correctly reproduces the numerical simulation as shown in Fig. \ref{Fig4}. In this regime, the switching rate (Eq. \eqref{gamma}) takes a much simpler form: 
\begin{equation}\label{gamma2}
\begin{gathered}
\Gamma=\frac{2  }{\pi} 
   \frac{\left| \Delta \right| }{G}\sqrt{G^2-\Delta ^2}\times\\
   \times\exp\left[-\frac{2\sqrt{G^2 -\Delta ^2 }}{\eta }+\frac{2 \Delta   \arctan\left(\frac{\sqrt{G^2 -\Delta ^2 }}{\Delta}\right)}{\eta}\right].
\end{gathered}
\end{equation}
where we use $G,\Delta>>\eta$.  When $\Delta>>\eta$, the exponential term in Eq. \eqref{gamma2} becomes dependent on $\Delta$. The first contribution to the exponent in Eq. \eqref{gamma2} is proportional to the separation between quantum states $|\alpha_{cl}|^2=\sqrt{G^2 -\Delta ^2 }/\eta$. As two states become closer to each other with increasing detuning, tunneling between them becomes faster. However, there is another contribution to the exponent in Eq. \eqref{gamma2} . It is proportional to $\Delta$ and arises from the rotation of $\alpha_{cl}$ in the complex plane with increasing detuning. The corresponding phase factor is as follows $e^{i2\theta_{cl}}=(\sqrt{G^2-\Delta ^2}+i \Delta)/G$. This new term significantly accelerates the growth of the switching rate \eqref{gamma3} with increasing detuning, as shown in (Fig. \ref{Fig4}). The coalesce of the quantum states takes place at the critical point $\Delta= G$, where the separation between the states $|\alpha_{cl}|$ becomes zero and a dissipative phase transition occurs \cite{Mylnikov2025, Mylnikov2022, Bartolo2016}. The switching rate \eqref{gamma2} has the following asymptotics near the critical point $\Delta- G<<G$:
\begin{equation} \label{GammaCP}
\begin{gathered}
\Gamma\approx\frac{2}{\pi} 
   \sqrt{2G(G-\Delta)}\exp\left[-\frac{4 \sqrt{2} (G-\Delta )^{3/2}}{3 \eta  \sqrt{G}}\right].
\end{gathered}
\end{equation}
As we show below, the switching rate \eqref{GammaCP} near the critical point can be estimated using simple and straightforward arguments within the Wigner representation.  

\subsection{The Wigner representation}

It is well known that the density matrix has several equivalent phase representations, which are different in ordering the operators \cite{Scully}. In a previous Section, we used a P-representation, which corresponds to a normal ordering of the operators, and has a Fokker-Planck-like equation of motion. However, one can use a symmetrically ordered Wigner representation instead of the P-representation. The main disadvantage of this approach is the absence of the Fokker-Planck-like equation of motion. The equation of motion for the Wigner function contains third-order derivatives, which significantly complicates any further calculations. However, an approximate equation with a Fokker-Planck-like structure can be obtained by truncating the third-order derivatives \cite{Mylnikov2025}. Nevertheless, a potential condition for an approximate Wigner equation is absent, and it is quite challenging to find its stationary solution. Fortunately, some progress can be made if the system stays near the critical point $\Delta\approx G$, where the dissipative phase transition occurs \cite{Downing2023}. In this regime, the stationary Wigner function has the following form \cite{Mylnikov2025}: 
\begin{equation}\label{Wigner}
\begin{gathered}
W(x,v)=\exp\left[-\frac{mv^2}{2T_{\mathrm{eff}}}-\frac{V(x)}{T_{\mathrm{eff}}}\right],
\end{gathered}
\end{equation}
where $x=\sqrt{2} \mathrm{Re}[\alpha]$ and $p=\sqrt{2} \mathrm{Im}[\alpha]$ are the photonic quadratures, $v=-2G p-\eta x^3/2$ is an auxiliary variable that plays a role of the effective velocity, $m=1/2G$ is the effective mass, $T_{\mathrm{eff}}=G/2$ is the effective temperature, and $V(x)$ is the effective potential:
\begin{equation}
V(x)=\frac{\Delta-G}{2}x^2+\frac{\eta^2}{48G}x^6,
\end{equation}
which is very similar to the thermodynamic potential in the Landau theory. Above the critical point $\Delta<G$, potential has two stable minima at $\pm x_0$, where $x_0=(8G(G-\Delta))^{1/4}$, and one unstable saddle point at $x=0$. For additional details on the calculation of the Wigner distribution function near the critical point, see also our previous paper \cite{Mylnikov2025}. The Wigner function \eqref{Wigner} is the same as the distribution function of a Brownian particle at thermal equilibrium with coordinate $x$ and velocity $v$. This comparison becomes clear if we write the Langevin equation on the $x$ photonic quadrature \cite{Mylnikov2025}:  
\begin{equation}\label{Langevin}
\p_t^2x+2\eta x^2\p_t x=-\frac{1}{m}\p_xV(x)+x\sqrt{\frac{4 \eta T}{m}}\xi,
\end{equation}
where $\xi(t)$ is a Gaussian white noise.  The effective Langevin Eq. \eqref{Langevin} is very similar to the equation of motion of a damped nonlinear classical oscillator in a multiplicative white-noise environment.
    In this Section, we will only discuss a qualitative idea of how to estimate the switching rate \eqref{GammaCP} using the Wigner representation. The main idea is that the effective potential $V(x)$ near the critical point has two minima separated by a barrier. At a finite effective temperature $T$, the particle can activate from one minimum to another through "thermal" activation. The probability of such transitions is determined by the Arrhenius law \cite{Hanggi1990}:
\begin{equation}\label{GammaW}
\Gamma=\frac{\omega_0}{2\pi} \exp(-|V(x_0)|/T_{\mathrm{eff}}),
\end{equation}
where $\omega_0=\sqrt{V''(0)/m}$ is the typical frequency near the saddle point, $V''(0)$ is the second derivative of the potential with respect to $x$. The expression \eqref{GammaW} follows from the classical Kramers theory. The calculated $\omega_0$ and $V(x_0)/T_{\mathrm{eff}}$ are given as follows:
\begin{equation}
\begin{gathered}
\omega_0=\sqrt{2G}\sqrt{G-\Delta},\\\quad V(x_0)/T_{\mathrm{eff}}=-\frac{4\sqrt{2}(G-\Delta)^{3/2}}{3\eta\sqrt{G}}.
\end{gathered}
\end{equation}
Substituting these expressions into the Kramer's formula \eqref{GammaW}, we reproduce the switching rate \eqref{GammaCP}. Thus, both the Wigner representation and the P-representation give the same answer near the critical point. 

\subsection{Nonzero Kerr nonlinearity and Kerr cat qubit}
\begin{figure}
\includegraphics[width=0.42\textwidth]{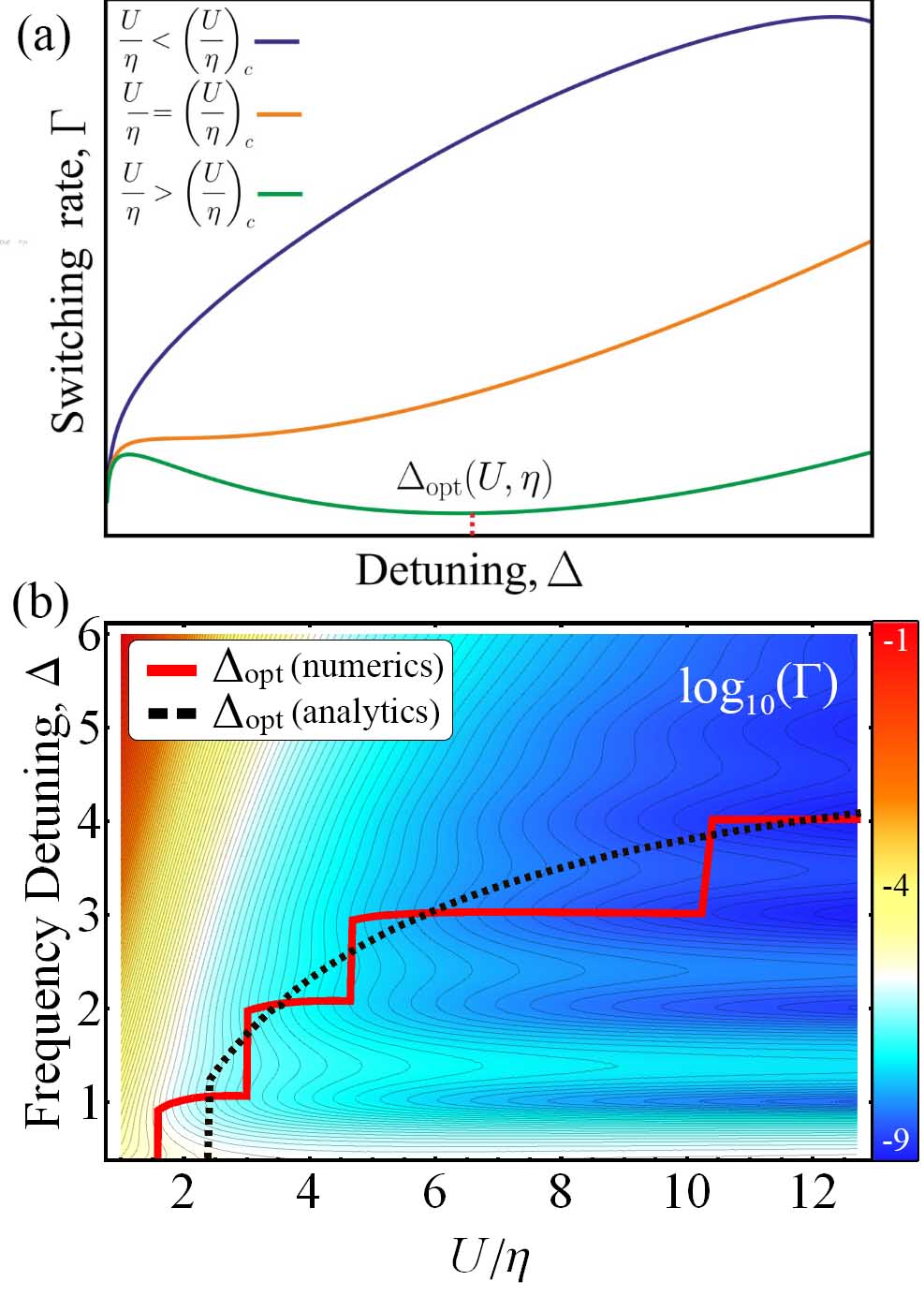}
\caption{(a) Schematic dependence of the switching rate $\Gamma$ on detuning $\Delta$, for different values of $U/\eta$. The minimum of the green curve corresponds to $ \Delta_{\rm {opt}}>0$. (b) The common logarithm of the switching rate, $\log_{10}\Gamma$, as a function of the frequency detuning $\Delta$ and the ratio $U/\eta$  is calculated using numerical diagonalization of the Liouvillian superoperator. The optimal detuning $\Delta_{\text{opt}}$ vs  $U/\eta$, calculated numerically (red curve) and via the analytical expression \eqref{gamma} (black dashed curve). The two-photon pump rate and modulus of the nonlinear coupling constant are set to $G = 6$, $|\kappa_2|=\sqrt{U^2+\eta^2}=1$, correspondingly.}
\label{Fig5}
\end{figure}

In the previous Sections, we discussed the switching rate between the two classical metastable states of a parametric oscillator with two-photon dissipation. Here, we will analyze how Kerr nonlinearity affects this phenomenon. In the resonance case $\Delta=0$, the switching rate vanishes. This is a consequence of a strong $Z_2$ symmetry that persists even for nonzero Kerr nonlinearity \cite{Kamenev, Thompson2022, Lieu2020}, ensuring the stability of the Kerr cat qubit. However, nonzero detuning results in a finite switching rate, as shown in Fig. \ref{Fig3}. As demonstrated earlier (Eq. \eqref{exactRate}), the switching rate for a dissipative cat qubit ($U=0$)  is exponentially small and its prefactor grows quadratically with frequency detuning. Remarkably, this behavior persists even for non-zero Kerr nonlinearity. After slight modifications of Eq. \eqref{exactRate} it is easy to guess the right answer for the switching rate in the small detuning regime ($\Delta<<\eta,U$): 
\begin{equation}\label{GammaSmallDetunings}
\begin{gathered}
\Gamma=\frac{4 \Delta^2 }{\sqrt{\eta^2+U^2}} 
 \exp\left[-2G/\sqrt{\eta^2+U^2}\right].
\end{gathered}
\end{equation}
This result agrees well with the numerical simulations, as shown in Fig. \ref{Fig3}. In the large detuning regime  ($\Delta>>\eta,U$), the switching rate changes asymptotically from Eq. \eqref{GammaSmallDetunings} to "potential barrier approximation" \eqref{gamma}. The comparison between analytical expression (Eq. \eqref{gamma}) and numerical simulation is shown in Fig. \ref{Fig4}. For simplicity, we introduce the following parametrization of the nonlinear coupling constants: $\eta=|\kappa_2|\cos(\theta)$ and $U=|\kappa_2|\sin(\theta)$, where $|\kappa_2|=\sqrt{\eta^2+U^2}$. Thus, we can explore the entire hybrid region between the dissipative limit ($\eta>>U$) and the Kerr limit ($U>>\eta$) by continuously varying $\theta$ from $0$ to $\pi/2$. For small Kerr nonlinearity, the switching rate increases monotonically with detuning. This means that the switching rate has a minimum value at zero detuning ($\Delta=0$). However, at large Kerr nonlinearities, significant changes occur. After some critical Kerr constant, the switching rate acquires a nonmonotonic dependence on frequency detuning $\Delta$ \cite{Gravina2023}. For our parameters ($G=6, |\kappa_2|=\sqrt{\eta^2+U^2}=1$), this happens for $ U/\eta\approx 1.6$. Let us define the critical ratio $(U/\eta)_c$ as the transition point from monotonic to non-monotonic behavior as shown in Fig. \ref{Fig5}a. Above this point ($U/\eta>(U/\eta)_c$), a minimum of the switching rate appears at non-zero detunings $\Delta_{\text{opt}}>0$ (green line in Fig.\ref{Fig5}a). This nonmonotonic behavior of the switching rate has an immediate practical application in the paradigm of the critical Schrödinger Cat Qubit, where both two-photon dissipation and Kerr nonlinearity are present and large frequency detuning provides optimal performance of the critical cat code \cite{Gravina2023}. We also need to mention the emergence of oscillations in the switching rate for small two-photon dissipation (bright orange curves in Fig. \ref{Fig4}). Local dips arise at $\Delta=mU$, where $m$ is a natural number, and originate from tunnelling-suppression effects \cite{Frattini2024, Venkatraman2024}. 

The minimization of the switching rate at finite detuning  $\Delta_{\text{opt}}>0$ is also illustrated in Fig. \ref{Fig5}b, where the phase diagram of the switching rate $\Gamma$ is considered as a function of $\Delta$ and $U/\eta$.  The dependence of the optimal detuning $\Delta_{\text{opt}}$ on the ratio of $U/\eta$ is also shown in Fig. \ref{Fig5}b. One can see that $\Delta_{\text{opt}}$ makes a discontinuous jump from zero to finite value $\Delta_{\text{opt}}\approx|\kappa_2|$ near some critical ratio $ (U/\eta)_c\approx 1.6$. Increasing $U/\eta$ also increases the value of $\Delta_{\text{opt}}$, as shown by the red curve in Fig. \ref{Fig5}b. It can be seen that $\Delta_{\text{opt}}$ calculated numerically exhibits steps that are a consequence of tunnelling-suppression effects \cite{Frattini2024, Venkatraman2024}. To compare numerical simulations and the "potential barrier approximation", we also calculated $\Delta_{\text{opt}}$ that minimizes the analytical switching rate \eqref{gamma} (black curve in Fig. \ref{Fig5}b). From Fig. \ref{Fig5}b it is evident that the "potential barrier approximation" gives a characteristic dependence of  $\Delta_{\text{opt}}$ on the ratio $U/\eta$, but it struggles to describe its "quantized" steps.  

\begin{figure}
\includegraphics[width=0.45\textwidth]{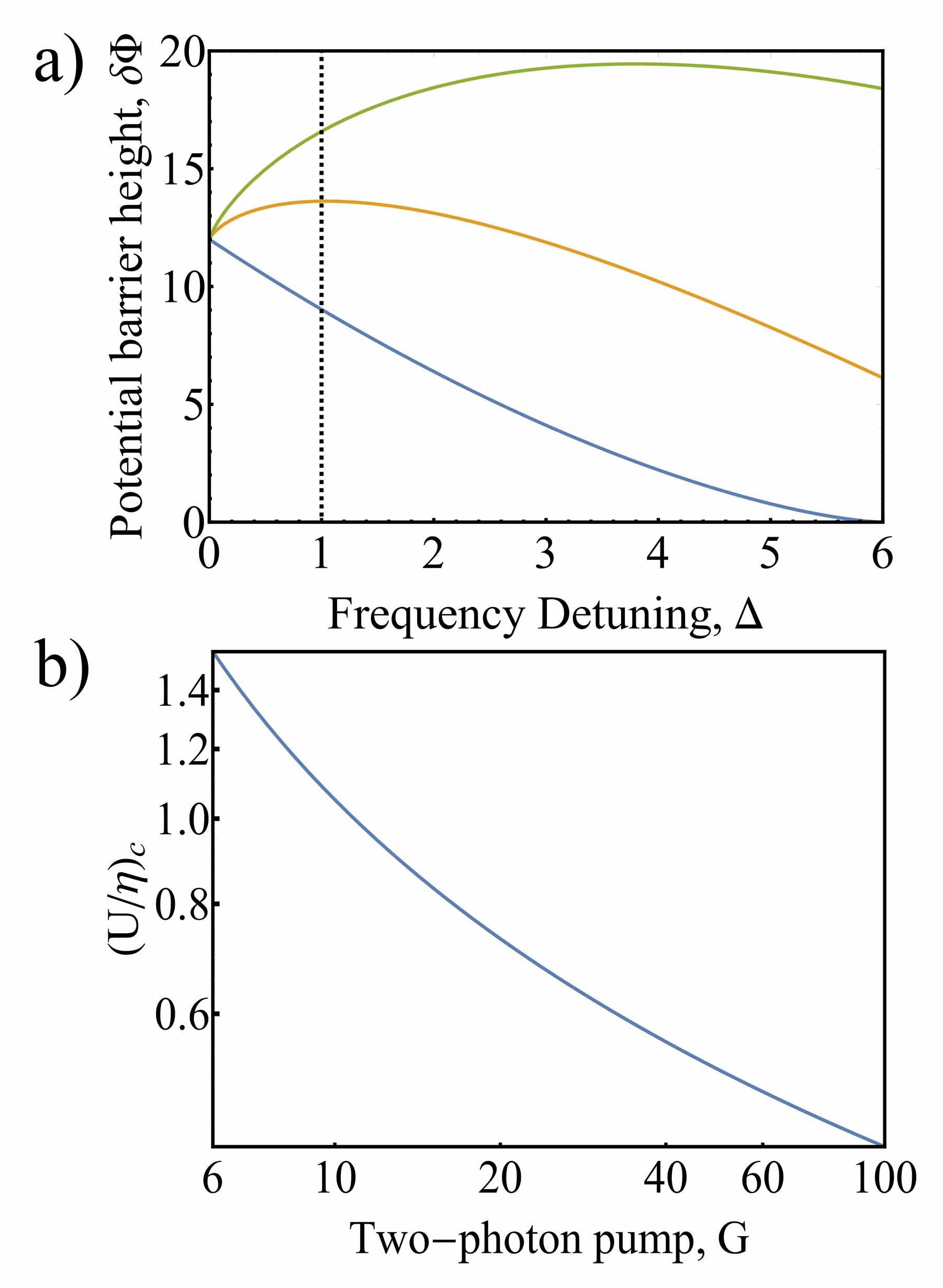}
\caption{(a) The potential barrier height \eqref{Porential} vs frequency detuning $\Delta$ for different values of the ratio $U/\eta=0$ (blue line), $U/\eta=1.4$ (orange line) and $U/\eta=6.3$ (green line). The two-photon pump rate and the modulus of the nonlinear coupling constant are set to $G = 6$, $|\kappa_2|=\sqrt{U^2+\eta^2}=1$, correspondingly. The dotted line indicates the detuning $\Delta=|\kappa_2|$, separating the region of small detunings ($\Delta<<|\kappa_2|$) from large detunings ($\Delta>>|\kappa_2|$), where the "potential barrier approximation" is valid. (b) The critical ratio $(U/\eta)_c$, defined in Eq. \eqref{CritRatio}, vs the two-photon pump rate $G$. We consider the modulus of the nonlinear coupling constant $|\kappa_2|=\sqrt{U^2+\eta^2}=1$.}
\label{Fig6}
\end{figure}

The existence of nonzero optimal detuning has a simple explanation within the P-representation theory. The switching rate \eqref{gamma} has the form of the Arrhenius equation, where the difference between the effective potentials at the classical and saddle points defines the "barrier energy" \eqref{Porential}. Quantum fluctuations can activate the system across the "barrier" from one metastable state to another. The higher the "barrier" the less the rate of switching between quantum states. As shown in Fig. \ref{Fig6}a, increasing the detuning lowers the "barrier" as the states move closer to each other in phase space for zero Kerr nonlinearity (blue curve in Fig. \ref{Fig6}a). But starting from some critical Kerr nonlinearity, this tendency is replaced by a growth in the "barrier" height with increasing detuning (orange curve in Fig. \ref{Fig6}a). For Kerr constants larger than the critical one, the maximum of the "potential barrier" is located at finite frequency detuning (green curve in Fig. \ref{Fig6}a).

These straightforward arguments can help us to estimate the critical ratio $(U/\eta)_c$, for which the switching rate becomes nonmonotonic for the first time. To do this, we calculate the first derivative of the effective potential  \eqref{Porential} at $\Delta=|\kappa_2|$, where the "potential barrier approximation" starts to be valid. The ratio $(U/\eta)_c$ at which the first derivative changes sign is critical. After some simple calculations, we found the following answer: 
\begin{equation}\label{CritRatio}
\begin{gathered}
(U/\eta)_c\approx\\
\tan\left(\frac{2}{\pi} \left(\sqrt{\ln ^2\left(\frac{G}{e|\kappa_2|}\right)+\frac{\pi
   ^2}{2}}-\ln \left(\frac{G}{e|\kappa_2|}\right)\right)\right),
\end{gathered}
\end{equation}
where $e$ is an Euler's number, and we assume that the modulus of the nonlinear coupling constant $|\kappa_2|=\sqrt{U^2+\eta^2}$  remains fixed. From Eq. \eqref{CritRatio} one can conclude that the critical ratio tends to zero slowly with increasing two-photon pump rate (Fig. \ref{Fig6}b) because it is inversely proportional to the logarithm $(U/\eta)_c\approx\pi/\ln \left(G^2/(e^2|\kappa_2|^2)\right)$. Thus, for states with a not very large two-photon pump ($G/|\kappa_2|\approx6-10$), the minimization of the switching rate at finite detuning begins in the hybrid regime ($U/\eta\approx1-2$). However, increasing the separation between states ($G/|\kappa_2|\approx10-100$) moves the boundary into the dissipative regime ($U/\eta<<1$). 

If we increase the ratio  $U/\eta$ above the critical one, the optimal detuning starts to increase, as shown in Fig. \ref{Fig5} (red and black curves). However, we need to note that our theory is limited to the region $|\Delta|<G$. This happens because the saddle point becomes stable at large detunings (see Appendix \eqref{AppA} for details), and the Kramer's theory needs to be modernized.

This imposes the following restrictions on the determination of the optimal detuning value: $U/\eta\lesssim(8 \pi/3)(G/|\kappa_2|)$. For these nonlinearities, the optimal detuning $\Delta_{\text{opt}}\approx G+|\kappa_2|$.  The minimal value of the switching rate for this case is as follows: 
\begin{equation}
\Gamma \approx\frac{2 G}{\pi }\exp(-4G/U),
\end{equation}
where we use $\eta<<U$. The exponential factor $4G/U$ is twice greater than the factor near zero detuning $2G/U$ as shown in Eq. \eqref{GammaSmallDetunings}. This significant suppression of the switching rate near the nonzero detuning can be beneficial for practical applications based on critical cat codes.

\section{Conclusion }

In this work, we apply the Kramer's theory to describe the dissipative gap of the Kerr cavity with nonzero detuning, two-photon dissipation, and driving. This system features two oscillatory states that form a qubit basis. However, this qubit is subject to decoherence with an exponentially small switching rate. We demonstrate that switching between metastable states due to quantum fluctuations can be modelled in the framework of the semiclassical activation process across a potential barrier. To establish this, we employ the P-distribution that obeys the Fokker–Planck–type master equation with a known stationary solution. The switching (or Bit-flip) rate was calculated analytically using the well known "potential barrier approximation". This result agrees well with the dissipative gap obtained from the numerical diagonalization of the Liouvillian superoperator.

In the dissipative limit ($U=0$), the switching rate increases with detuning because the two states move closer in the phase space, accelerating tunnelling. However, the exponential factor in the switching rate exhibits a complex dependence on system characteristics, contrary to the naive assumption that it would be proportional only to the average photon number. We also compared our results with those of the Wigner representation, showing identical behavior near the dissipative phase transition point $\Delta\approx G$. 

For nonzero Kerr nonlinearity ($U\neq0$), the switching rate becomes a nonmonotonic function of frequency detuning reaching a minimum at a finite value $\Delta_{\text{opt}}>0$. We explain this effect by analyzing the semiclassical activation process across a potential barrier. The effect of detuning on this barrier, however, depends critically on the Kerr nonlinearity: while increased detuning lowers the barrier for low or zero nonlinearity, beyond a critical Kerr nonlinearity, it instead causes the barrier height to grow, shifting its maximum to a finite detuning. The obtained results are of practical importance for the implementation of stable critical dissipative cat qubits and for optimizing the parameters of two-photon resonators in quantum computing architectures.

\begin{acknowledgments}
G.S. Sokolovskii thanks the Russian Science Foundation (Project No. 21-72-30020-P) for financial support of numerical simulations. V. Yu. Mylnikov acknowledges the support of the theoretical study by the Foundation for the Advancement of Theoretical Physics and Mathematics “BASIS.” We acknowledge fruitful discussions with Alex Kamenev, Igor S. Burmistrov and Charles A. Downing.
\end{acknowledgments}

\appendix
\section{THE STABILITY OF STATIONARY SOLUTIONS IN THE SEMICLASSICAL APPROXIMATION
}
\label{AppA}
In this appendix, we examine the stability of stationary solutions of Eq. \eqref{alpha}:
\begin{equation}\label{A1}
    \begin{aligned}
        \frac{d\alpha}{dt} &= A_{\alpha},\quad \frac{d\beta}{dt} = A_{\beta}.
    \end{aligned}
\end{equation}
Here, we introduce the auxiliary variable $\beta=\alpha^*$ and denote: 
\begin{equation}
 \begin{split}
    & A_{\alpha} = i\Delta \alpha + G\beta - (\eta + iU)\alpha^{2}\beta,\\
    &A_{\beta} =-i\Delta \beta+G\alpha-(\eta-i U)\beta^2\alpha.
 \end{split}
\end{equation}
To determine the stability of the stationary points, we construct the Jacobian matrix whose elements are obtained by differentiating the right-hand sides of the Eq. \eqref{A1} with respect to the corresponding variables:
\begin{equation}
\begin{aligned}
&J(\alpha,\beta) = \left(
\begin{array}{cc}
 \partial_{\alpha}A_{\alpha} & \partial_{\beta}A_{\alpha}\\
 \partial_{\alpha}A_{\beta} & \partial_{\beta}A_{\beta}
\end{array}
\right)=\\
&= \left(
\begin{array}{cc}
 i \Delta -2 \alpha  \beta  (\eta +i U) & G-\alpha ^2 (\eta +i U) \\
 G-\beta ^2 (\eta -i U) & -i \Delta -2 \alpha  \beta  (\eta -i U) \\
\end{array}
\right)
\end{aligned}
\end{equation}
\begin{figure}
\includegraphics[width=0.45\textwidth]{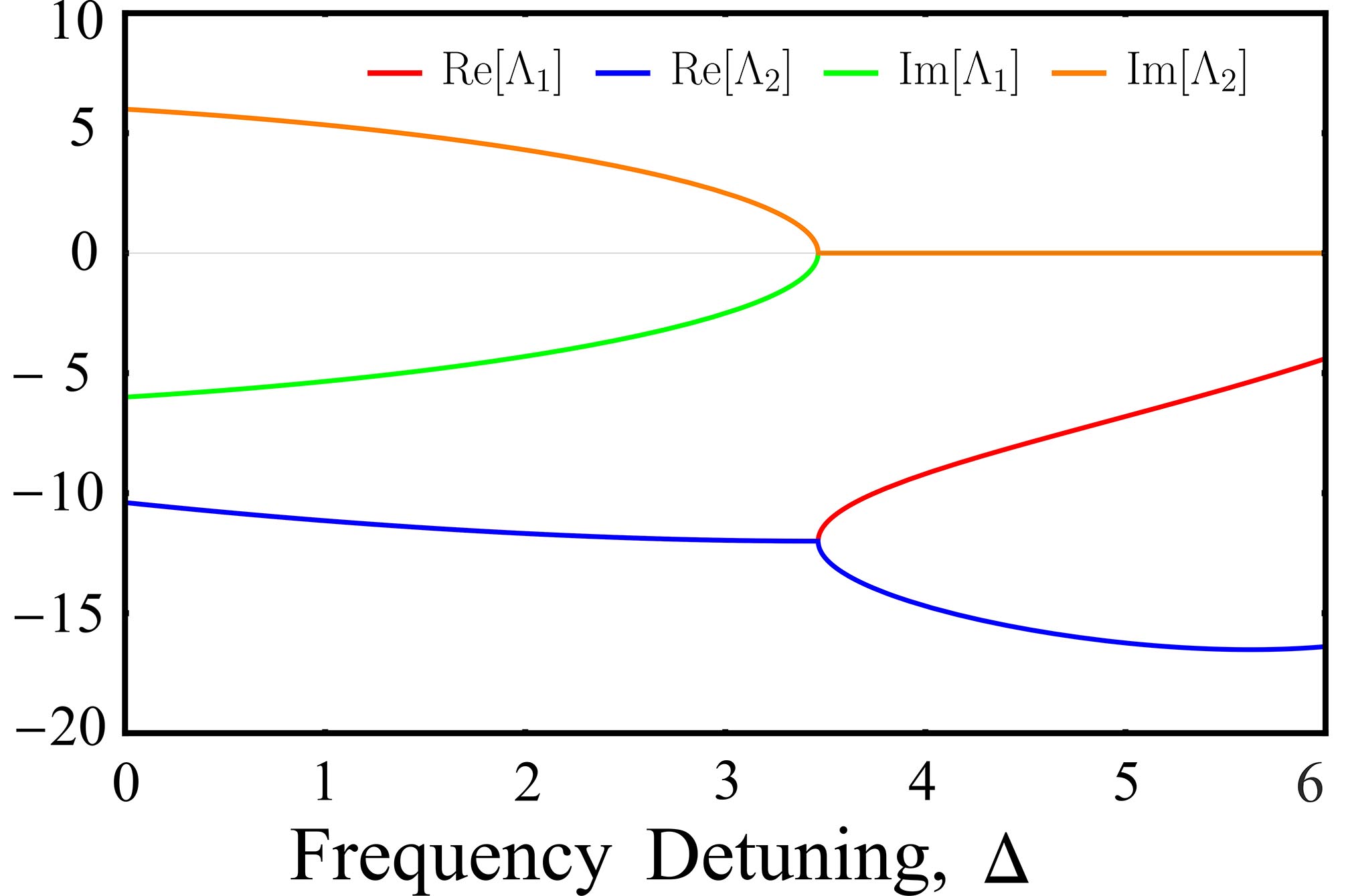}
\caption{The eigenvalues $\Lambda_{1,2}$ correspond to nontrivial stationary point $(\alpha=\alpha_0,\beta=\alpha^*_0)$ and are determined by Eq.~\eqref{EV12} vs the frequency detuning $\Delta$. The type of stability changes from a stable focus to a stable node at $\operatorname{Im}{\Lambda_{1,2}}=0$. The parameters are set to $G=6$, $U=1/2$, $\eta=\sqrt{3}/2$.}
\label{FigLambda12}
\end{figure}

\begin{figure}
\includegraphics[width=0.45\textwidth]{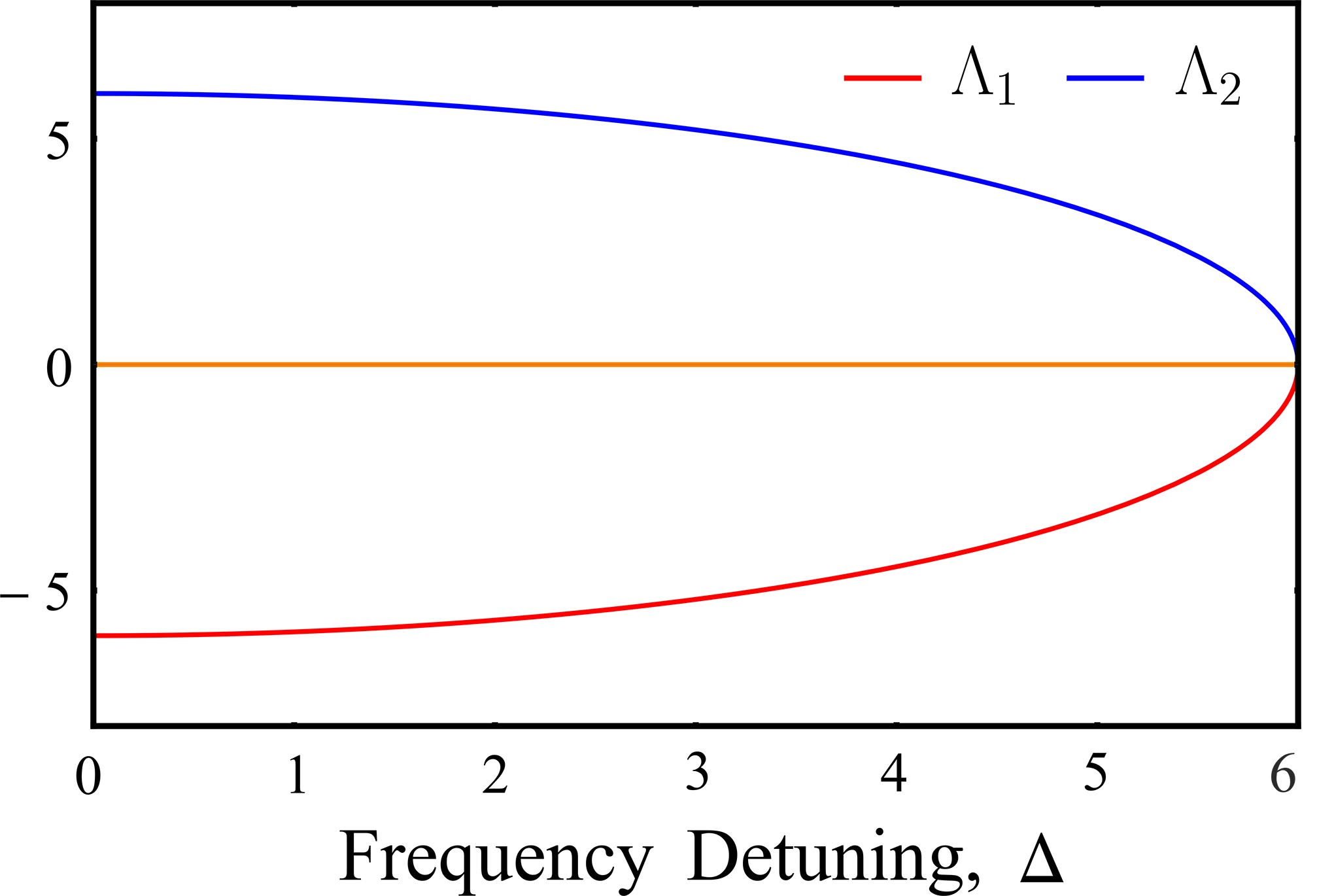}
\caption{
The eigenvalues $\Lambda_{1,2}$ correspond to trivial stationary point $(\alpha=0,\beta=0)$ and are determined by Eq.~\eqref{LambdaSad} vs the frequency detuning $\Delta$. For $\Delta < G$, the system exhibits a saddle point, whereas for $\Delta > G$, the equilibrium is a stable center. The parameters are set to $G=6$, $U=1/2$, $\eta=\sqrt{3}/2$.}
\label{FigLambdaSad}
\end{figure}
First, we consider the behavior near the nontrivial solution $(\alpha=\alpha_0,\beta=\alpha^*_0)$. The Jacobian matrix in this case takes the form:
\begin{equation}
\begin{aligned}\label{a0b0}
    &J(\alpha=\alpha_0,\beta=\alpha^*_0)= \\
    &\begin{pmatrix}
     i \Delta -2 n_0(\eta+i U) & \frac{\Delta}{G}\left(\Delta-n_0(U+i\eta)\right) \\
     \frac{\Delta}{G}\left(\Delta-n_0(U-i\eta)\right) & -i \Delta -2 n_0(\eta-i U)
    \end{pmatrix},
\end{aligned}
\end{equation}
For the matrix \eqref{a0b0}, the eigenvalues are:
\begin{align}\label{EV12}
    &\Lambda_{1,2}=-2n_0\left(\eta \pm i U\sqrt{1-\Delta/(Un_0)}\right)
\end{align}
First, consider the case where the expression under the square root in \eqref{EV12} is positive. Then the eigenvalues $\Lambda_{1,2}$ form a complex-conjugate pair. For our system,  $\operatorname{Re}\Lambda_{1,2}< 0$, which corresponds to a stable (attracting) focus.  Now consider the case where the expression under the square root in Eq.~\eqref{EV12} is negative. Then the eigenvalues are real. For our system, both eigenvalues are negative ($\Lambda_1 < 0$, $\Lambda_2 < 0$), which corresponds to a stable node. The type of stability changes from a stable node to a stable focus at the point $\Delta = \pm G (U/\eta$), as shown in Fig. ~\eqref{FigLambda12}. 

Next, consider the trivial solution $(\alpha=0,\beta=0)$. The Jacobian matrix then takes the form:
\begin{equation}\label{LambdaSad}
\begin{aligned}
    &J(\alpha=0,\beta=0)= \begin{pmatrix}
     i \Delta  & G \\
    G & -i \Delta
    \end{pmatrix}.
\end{aligned}
\end{equation}
The eigenvalues of matrix \eqref{LambdaSad} are:
\begin{equation}\label{LambdaSad}
    \Lambda_{1,2}=\pm\sqrt{G^2-\Delta^2}.
\end{equation}
Now consider the case $|\Delta|<G$, where the eigenvalues are real and have opposite signs. In this situation, one eigenvalue is positive and the other is negative, so the stationary solution is a saddle point and, therefore, unstable. The phase trajectories diverge along the direction of the positive eigenvector and converge along the negative one. If $|\Delta|>G$, the eigenvalues are purely imaginary, $\Lambda_{1,2} = \pm i \sqrt{\Delta^2 - G^2}$. In this case, the equilibrium is a stable center, and the phase trajectories are closed, describing oscillations around the stationary solution. These facts are illustrated in Fig.~\ref{FigLambdaSad}.

\section{CALCULATION OF THE DECOHERENCE RATE }
\label{AppB}
This appendix is devoted to the calculation of the decoherence rate \eqref{Kramers1}. Let us start with the effective potential \eqref{PotentialP}. At a saddle point $(\alpha=0,\beta=0)$, it is given by: 
\begin{equation}\label{pot0}
\begin{gathered}
\Phi(0,0)=\left(1+\frac{\Delta  U}{\eta ^2+U^2}\right) \ln
   \left(\frac{G^2}{\eta ^2+U^2}\right)+\\+\frac{2 \Delta 
   \eta }{\eta
   ^2+U^2}\arctan\left(\frac{U}{\eta }\right).
\end{gathered}
\end{equation}
The potential at the classical point $(\alpha=\alpha_{cl},\beta=\alpha_{cl}^*)$ is more complicated and includes three contributions: 
\begin{equation} \label{potcl}
\Phi(\alpha_{cl},\alpha_{cl}^*)=\Phi_1+\Phi_2+\Phi_3,
\end{equation}
where $\Phi_1$ is the first contribution to the potential \eqref{Kramers1} arising from $-2\alpha_{cl}\alpha_{cl}^*$:
\begin{equation} 
\Phi_1=-\frac{2 \left(\sqrt{G^2 \left(\eta ^2+U^2\right)-\Delta ^2 \eta^2}+\Delta  U\right)}{\eta ^2+U^2}-2.
\end{equation}
The second and third contributions ($\Phi_2$ and  $\Phi_3$) come from the logarithm at \eqref{Kramers1} and are determined by the argument of $\alpha_{cl}$ \eqref{angles1} :
\begin{equation} 
\begin{gathered}
\Phi_2=\frac{2 \Delta  \eta   \arctan\left(\frac{\sqrt{G^2 \left(\eta
   ^2+U^2\right)-\Delta ^2 \eta ^2}}{\Delta  \eta }\right)}{\eta ^2+U^2}+\\
  +\frac{4 \Delta  \eta   \arctan\left(\frac{U}{\eta }\right)}{\eta ^2+U^2} 
  +\frac{2 \Delta  \eta   \arctan\left(\frac{\eta }{\Delta +U}\right)}{\eta ^2+U^2},
\end{gathered}
\end{equation}
\begin{equation} 
\Phi_3=\left(1+\frac{\Delta  U}{\eta ^2+U^2}\right) \ln \left(\frac{\eta ^2+(\Delta
   +U)^2}{\eta ^2+U^2}\right).
\end{equation}
The next step is a calulation of the Hessian at both the classical $(\alpha=\alpha_{cl},\beta=\alpha_{cl}^*)$ and saddle $(\alpha=0,\beta=0)$ points. Firstly, we evaluate the Hessian at the arbitrary point of the four-dimensional complex plane $(\alpha,\beta)$:
\begin{equation}\label{Hessian} 
\begin{gathered}
\nabla^2\Phi(\alpha,\beta)=\left(
\begin{array}{cc}
 \frac{\partial^2\Phi}{\partial\alpha^2} &  \frac{\partial^2\Phi}{\partial\alpha\partial\beta} \\
\frac{\partial^2\Phi}{\partial\alpha\partial\beta} &  \frac{\partial^2\Phi}{\partial\beta^2} \\
\end{array}
\right)=\\=\left(
\begin{array}{cc}
 -\frac{2 i (\Delta -i \kappa_2 ) \left(G+\kappa_2 \alpha ^2\right)}{\left(G-\kappa_2\alpha
   ^2 \right)^2} & -2 \\
 -2 & \frac{2 i (\Delta +i \kappa_2^*) \left(G+\kappa_2^* \beta
   ^2\right)}{\left(G-\kappa_2^*\beta ^2\right)^2} \\
\end{array}
\right),
\end{gathered}
\end{equation}
where $\kappa_2=\eta+i U$. At the saddle point \eqref{Hessian} has the following form:
\begin{equation}\label{Hess0}
\begin{gathered}
\nabla^2\Phi(0,0)=\left(
\begin{array}{cc}
 -\frac{2 i (\Delta -i \kappa_2 )}{G} & -2 \\
 -2 & \frac{2 i (\Delta +i \kappa_2^*) }{G} \\
\end{array}
\right),
\end{gathered}
\end{equation}
with the following eigenvalues:
\begin{equation}\label{eigen1}
\Lambda_1(0,0)=-\frac{2 \left(\sqrt{G^2-(\Delta +U)^2}+\eta\right) }{G},
\end{equation}
\begin{equation} 
\Lambda_2(0,0)=\frac{2 \left(\sqrt{G^2-(\Delta +U)^2}-\eta\right) }{G},
\end{equation}
and determinant: 
\begin{equation}\label{det0}
\det(\nabla^2\Phi(0,0))=-\frac{4 \left(G^2-(\Delta +U)^2-\eta ^2\right)}{G^2}.
\end{equation}
At the classical point, the deteminant of the Hessian \eqref{Hessian} is given by:
\begin{equation}\label{detcl}
\begin{gathered}
\det(\nabla^2\Phi(\alpha_{cl},\alpha_{cl}^*))=\frac{16 \sqrt{G^2 \left(\eta ^2+U^2\right)-\Delta ^2
   \eta ^2}}{\eta ^2+U^2}\\
\times\left(\frac{\sqrt{G^2 \left(\eta ^2+U^2\right)-\Delta
   ^2 \eta ^2}+\Delta  U +U^2+\eta ^2}{\eta ^2+(\Delta +U)^2}\right).
   \end{gathered}
\end{equation}
Using equations \eqref{pot0} ,\eqref{potcl}, \eqref{eigen1},\eqref{det0} and \eqref{detcl} one can directly obtain the decoherence rate \eqref{Kramers1}. However, as long as the nonlinearities ($U$ and $\eta$) are relatively small compared to the two-photon pump rate $G$ one can approximate part of the preexponential factor in the following way:  

\begin{equation}\label{prefgol}
\begin{gathered}
D_0\frac{ |\Lambda_1(0,0)|}{2 \pi}\sqrt{\frac{\det(\nabla^2\Phi(\alpha_{cl},\alpha_{cl}^*))}{|\det(\nabla^2\Phi(0,0))|}}\approx\\
\approx
\frac{2}{\pi }
\left(\frac{\eta ^2+(\Delta +U)^2}{G^2}\right)^{-1/2}\times\\
\left(G^2 \left(\eta ^2+U^2\right)-\Delta ^2 \eta ^2\right)^{1/4}\times\\
\sqrt{\frac{\sqrt{G^2 \left(\eta ^2+U^2\right)-\Delta ^2
   \eta ^2}+\Delta  U}{\eta ^2+U^2}}.
   \end{gathered}
\end{equation}
We have demonstrated in the Section IV.B-D, that the "potential barrier approximation" is valid only for large frequency detunings ($\Delta>>\eta, U$). Thus, we can simplify Eqs. \eqref{pot0}-\eqref{prefgol}. Firstly, one can approximate the exponent with effective potential at the classical and saddle points:  
\begin{equation}\label{approxPot}
e^{\Phi(\alpha_{cl},\alpha_{cl}^*)-\Phi(0,0)}\approx
(\Delta/G)^2e^{-\delta\Phi},
\end{equation}
where $\delta\Phi$ is an effective potential barrier height and it is given by Eq. \eqref{Porential}. It should be noted that the same effective potential barrier $\delta\Phi$ can be found from the instanton approach \cite{Carde2025}. We also need to approximate the preexponential factor \eqref{prefgol}: 
\begin{equation}\label{approxPref}
\begin{gathered}
D_0\frac{ |\Lambda_1(0,0)|}{2 \pi}\sqrt{\frac{\det(\nabla^2\Phi(\alpha_{cl},\alpha_{cl}^*))}{|\det(\nabla^2\Phi(0,0))|}}\approx
\\\approx\frac{2}{\pi }
\frac{G }{|\Delta|}
\left(G^2 \left(\eta ^2+U^2\right)-\Delta ^2 \eta ^2\right)^{1/4}\times\\
\sqrt{\frac{\sqrt{G^2 \left(\eta ^2+U^2\right)-\Delta ^2
   \eta ^2}+\Delta  U}{\eta ^2+U^2}}.
\end{gathered}
\end{equation}
After substitution of the Eq. \eqref{approxPot} and \eqref{approxPref} into the switching rate \eqref{Kramers1} we obtain the final expression \eqref{gamma}.
\appendix

\bibliography{main}

\end{document}